\documentstyle[12pt]{article}

\begin{document}
\newcommand{\be}{\begin{equation}}
\newcommand{\ee}{\end{equation}}
\newcommand{\al}{\alpha}
\title
{\bf 
\begin{flushright}
{\large \bf TTP98-01\\[-5mm]
January 1998}\\[15mm]
\end{flushright}
Coulomb resummation
for $b \bar b$ system near threshold and precision
determination of $\al_s$ and $m_b$.}

\author{
 {\bf J.H.K{\"u}hn}, \\
  {\small {\em Institut f{\"u}r Theoretische Teilchenphysik
  Universit{\"a}t Karlsruhe}}\\
  {\small {\em D-76128 Karlsruhe, Germany}}\\[3mm]
  {\bf A.A.Penin and A.A.Pivovarov}\\
  {\small {\em Institute for Nuclear Research of the
  Russian Academy of Sciences,}}\\
  {\small {\em 60th October Anniversary
  Pr., 7a, Moscow 117312, Russia}}
  }

\date{}

\maketitle

\begin{abstract}
We analyze sum rules for
the $\Upsilon$ system with resummation of threshold effects on the
basis of the nonrelativistic Coulomb approximation.
We find for the pole mass of the bottom quark $m_b=4.75\pm 0.04~{\rm GeV}$
and for the strong coupling constant $\al_s(M_Z)=0.118\pm 0.006$.
The origin of the contradiction between two recent estimates
obtained within the same formal framework is clarified.
\end{abstract}

\thispagestyle{empty}

\newpage

\section{Introduction}

\noindent The knowledge of precise numerical
values of the coupling constant and quark masses
is important for obtaining accurate predictions within
QCD. The technique of sum rules
is extensively used both for the determination 
of these parameters from experimental data on low lying 
resonances and 
for the calculation of properties of higher excited states
\cite{NSVZ,KrasnPiv}.
The $\Upsilon$ system is a rather clean place
to obtain the values of the strong coupling constant $\al_s$ and also
of the $b$-quark pole mass $m_b$
which is a key parameter for the rich B-meson
physics. In sum rules for the $\Upsilon$ system there are some subtleties
that make them interesting by themselves, the main one being the
importance of Coulomb effects and the applicability of
the nonrelativistic
approximation as a rather good starting point for
the expansion in the coupling constant
$\al_s$ and the $b$-quark velocity $v$ which are both
small parameters of the problem \cite{Vol1}.
The expressions for the first few moments  of the spectral density 
are now available 
in ordinary perturbation theory (PT) 
with $\al_s^2$ accuracy \cite{Chet}, however,  they cannot
be used in theoretical formulas for sum rules directly
because the spectrum is well known
experimentally only for energies close
to threshold due to existence of sharp resonances while the
contribution of the continuum to these low moments is large in
comparison
with the resonance contribution.
For higher moments that are saturated by these low energy resonances
with better accuracy the ordinary PT expansion
in the coupling constant
is not applicable (the coefficients of the expansion grow fast with the
order of the moment) and a  resummation is required.
In this paper we analyze  the sum rules for the $\Upsilon$ system
with resummation of Coulomb effects
that are largely responsible for the growth of PT coefficients
to extract the values of $\al_s$ and $m_b$.
We consistently take into account the first order corrections
in $\al_s$ and give an estimate of the magnitude of the high
order corrections.

The analysis is based on the same formal approach
which has  already been applied in refs.~\cite{Vol2,PichJam}
along different lines
and with contradictory results.
In ref.~\cite{Vol2} the next-to-leading
approximation in $\al_s$ both to the high energy part of the correlator and
to the Coulomb potential was used
while in ref.~\cite{PichJam} further terms of the perturbative expansion
in $\al_s$ have been
partly included as well.
This prevents the  direct comparison of the results of those two papers
though both  claim
that the corrections are small and under control.
However, the
numerical results do not overlap within the respective error bars
and the clarification of the origin of this
difference is in order.
We find that
the difference is to large extent due to neglecting in ref.~\cite{PichJam}
the contribution of Coulomb
poles to the  moments.
We demonstrate that after including the contribution of the
Coulomb poles the results of two papers can be made completely
compatible by the corresponding choice of the scale in the
nonrelativistic Coulomb potential -- the low one ($\mu \sim 1~{\rm GeV}$)
for the case of  ref.~\cite{Vol2}
and of order of the $b$-quark mass for  ref.~\cite{PichJam}.

The result of ref.~\cite{Vol2} for the value of $\al_s(M_Z)$
is furthermore in  contradiction with the LEP data \cite{PDG}.
The main uncertainty enters the theoretical predictions through
the higher order $\al_s$ corrections
which become uncontrollable at the low normalization point
used in ref.~\cite{Vol2}. The uncertainties of the predictions
given in ref.~\cite{Vol2} are therefore strongly underestimated 
explaning thus
the discrepancy with the LEP value for $\al_s(M_Z)$.

The paper is organized as follows.
The stage is set in Sect.~2 where  
our notations are fixed. 
The numerical analysis of sum rules is done in Sect.~3 where 
the procedure of optimization for the nonrelativistic Coulomb Green function 
is described. In Sect.~4 we discuss the difference between the present 
analysis 
and those of refs. \cite{Vol2,PichJam}. Sect.~5 contains our
conclusions. Some explicit formulas are collected in the Appendix.  

\section{Sum rules and nonrelativistic approximation}

\noindent We perform our analysis by computing the moments directly
instead of using a representation for
them through the propagation function
of the nonrelativistic approximation in Euclidean time as 
done in ref.~\cite{Vol2}. The difference between these two
approaches is briefly discussed later.

The moments of the spectral density are defined to be dimensionless
and are given by the following expression
\[
{\cal M}_n= 
\left.{12\pi^2\over n!}(4m_b^2)^n{d^n\over ds^n}\Pi(s)\right|_{s=0}=
(4m_b^2)^n\int_0^\infty{R(s)ds\over s^{n+1}}
\]
where 
\[
R(s)=12\pi {\rm Im}\Pi(s+i\epsilon)
\]
and $\Pi(s)$ is the polarization function 
of the $b$-quark vector current
\[
\left(q_\mu q_\nu-g_{\mu \nu}q^2\right)\Pi(q^2)=
i\int dxe^{iqx}\langle 0|Tj_{\mu}(x)j_{\nu}(0)|0\rangle ,
\]
$j_\mu=\bar b\gamma_\mu b$.
The assumption of quark-hadron duality relates 
the imaginary part of the polarization
function $R(s)$ to the
experimentally measurable quantity
$R_b(s) = \sigma(e^+e^-\rightarrow {\rm hadrons}_{\,b\bar b})/
\sigma(e^+e^-\rightarrow \mu^+\mu^-)$
by the sum rules 
\[
{\cal M}_n= (4m_b^2)^n\int_0^\infty{R(s)ds\over s^{n+1}}
={(4m_b^2)^n\over Q_b^2}\int_0^\infty{R_b(s)ds\over s^{n+1}}
\]
with $Q_b=-1/3$.

The experimental values
of the moments
are obtained by using information on
the first six $\Upsilon$ resonances. Their leptonic widths
$\Gamma_{k}$ and masses $M_{k}$, $(k=1\ldots 6)$
are known rather precisely \cite{PDG}
\be
{\cal M}_n={(4m_b^2)^n\over Q_b^2}\left({9\pi\over \al_{QED}^2(m_b)}
\sum_{k=1}^6{\Gamma_{k}\over M_{k}^{2n+1}}
+\int_{s_0}^\infty\!{\rm d}s {R_b(s)\over s^{n+1}}\right)
\label{exmom}
\ee
while the last term is a continuum contribution.
Beyond the resonance
region, for energies larger than
$s_0\approx (11.2~{\rm GeV})^2$,
the spectral density $R_b(s)$ can be approximated by
the ordinary PT expression for the theoretical 
spectral density $R(s)$. The influence of the continuum on high moments
is almost negligible numerically 
and in any case under strict control.
Here $\al_{QED}^2(m_b)=1.07 \al^2$ \cite{PDG}.
The theoretical expression of the polarization function
in the nonrelativistic approximation that is justified for
the energy region near the threshold is given by \cite{Braun}
\be
\Pi(s)
=\left(1-4 C_F{\al_s\over \pi}\right){3\over 2 m_b^2}
G(0,0,k)
\equiv N_{pt}{3\over 2 m_b^2}
G(0,0,k)
\label{thmom}
\ee
where $C_F=4/3$. The energy of the nonrelativistic Coulomb $b\bar b$
system counted 
from the threshold is
$E=-k^2/m_b$, and there are several ways of 
introducing $k$ through the
relativistic parameter of energy square
$s$
\be
k = \left\{
\begin{array}{l}
{1\over 2}\sqrt{4m_b^2-s}\\
m_b\sqrt{-1+4 m_b^2/s}\\
\sqrt{m_b(-\sqrt{s}+2m_b)}
\end{array}\right.
\label{var}
\ee
which all are equivalent in
nonrelativistic approximation near the threshold. 
For $E<0$ ($s<4m_b^2$) the parameter $k$ is real 
while for $E>0$ ($s>4m_b^2$) the parameter $k$ 
is purely imaginary. The quantity
$G(0,0,E)$ is the nonrelativistic Green
function (GF) of the $b\bar b$ system that can be found exactly for Coulomb
potential  which is supposed to dominate the whole QCD interaction in the
energy region of interest. 
The $\al_s(\mu)$ correction in
eq.~(\ref{thmom}) (so called perturbative factor $N_{pt}$) 
reflects the contribution of high energies 
to the polarization function and has as
natural scale $\mu \sim m_b$.
We use the $\overline{\rm MS}$ scheme of
renormalization throughout the paper
and $m_b$ denotes the  pole mass of the $b$ quark, as already
mentioned in the Introduction.
The PT expression for the nonrelativistic potential
is known up to the
second order in $\al_s$ \cite{Peter}
\be
V=V_C+\Delta V=V_C+{\al_s\over 4\pi}V_1+
\left({\al_s\over 4\pi}\right)^2V_2
\label{potcor}
\ee
where $V_C$ is the Coulomb potential, $V_C=-C_F\al_s/ r$
and explicit expressions for
$V_{1,2}$
are given in the Appendix (eq.~(\ref{potcorr1})).
The corrections to the Coulomb potential
in eq.~(\ref{potcor})
generate corrections to the Coulomb GF $G_C$.
The full nonrelativistic GF in condensed notation (see Appendix for
more details) is given by q
\be
G=G_C+\Delta G=G_C-G_C\otimes\Delta V\otimes G_C=G_C+G_1+G_2.
\label{gfcor}
\ee
where $G_C$ is a Coulomb GF and $G_1$ and $G_2$ correspond to
$V_1$ and $V_2$ contributions (see eqs.~(\ref{G0},~\ref{G1},~\ref{G2})
of Appendix).

In principle one can develop a regular expansion for the polarization
operator near the threshold using Bethe-Salpeter equation
for four-point amplitude using Barbieri-Remiddi exact solution
\cite{BethSal}
as a
starting point.
This is however a rather difficult technical problem.
Meanwhile up to the first order in $\al_s$
eqs.~(\ref{thmom},~\ref{gfcor})
give a complete result. The self-consistent calculation of the
expansion of the polarization function
beyond the first order is not available,
so we will use the $G_2$  term of  eq.~(\ref{gfcor})
not for the regular computation but rather
for estimating the typical high order $\al_s$ contributions to the
sum rules.

\section{Numerical analysis of sum rules}
\noindent
We consider two methods to compute the nonrelativistic GF when higher
order perturbative corrections to the potential are included.
The first one is the direct 
computation when the splitting into the leading
term (Coulomb) and corrections is determined by the ordering in the 
$\overline{\rm MS}$ scheme, i.e. we count powers of $\al_s$.
This method in fact fails because the  expansion turns out 
to be badly divergent. 
The second methods is the optimized computation when the
splitting of the whole potential into the leading term (Coulomb) and
corrections is determined by the physics of the problem rather than
the artificial $\overline{\rm MS}$ charge. This technique gives solid
results when all expansions (for the effective charge and for the
nonrelativistic GF) converge well for the moments that are computed.

\subsection{Direct computation}
\noindent
To compute the theoretical part of sum rules 
we use the approximation (\ref{thmom}) for the polarization function.
As a function of the nonrelativistic variable  $E$ 
the exact nonrelativistic Coulomb Green function of 
the $b\bar b$ system $G(0,0,E)$ has its poles below threshold 
for real $E<0$  ($s<4m_b^2$) corresponding
to the  bound states, and a cut above the threshold at real
$E>0$
($s>4m_b^2$). 
Accordingly, 
the theoretical moments are
split  into
\[
{\cal M}_n={\cal P}_n+{\cal C}_n
\]
where ${\cal P}_n$ is the pole contribution from the
energy region below the threshold
\[
{\cal P}_n= (4m_b^2)^n\int_0^{4m_b^2}{R(s)ds\over s^{n+1}}
\]
and ${\cal C}_n$ is the  continuum contribution from the
energy region above the threshold
\[
{\cal C}_n= 
(4m_b^2)^n\int_{4m_b^2}^\infty{R(s)ds\over s^{n+1}}.
\]

\begin{table}[htb]
\begin{center}
\begin{tabular}{|c|c|c|c|} \hline
$n$& ${\cal P}_n^C$ & ${\cal P}_n^1$ & ${\cal P}_n^2$ \\ \hline
10 & 0.16017  & 0.23375  & 0.25756  \\
12 & 0.16631  & 0.25180  & 0.27743  \\
14 & 0.17271  & 0.27101  & 0.29859  \\
16 & 0.17939  & 0.29145  & 0.32111  \\
18 & 0.18635  & 0.31319  & 0.34507  \\
20 & 0.19360  & 0.33630  & 0.37057  \\ \hline
\end{tabular}
\end{center}
\caption{ The 0th, 1st and 2nd order pole
contributions to the moments}
\label{tab2}
\end{table}

Then the total moments are represented by the sum
\be
{\cal M}_n={\cal M}_n^C+{\cal M}_n^1+{\cal M}_n^2
\label{mseries}
\ee
where ${\cal M}_n^C$ is the contribution of a pure Coulomb
part,
${\cal M}_n^i$ are the contributions due to $G_i$
terms in eq.~(\ref{gfcor}).
The same representation is used for ${\cal P}_n$ and ${\cal C}_n$ parts
separately and also for $\Pi(s)$ and $R(s)$,
for instance, ${\cal P}_n={\cal P}_n^C+{\cal P}_n^1+{\cal P}_n^2$ and
so on.
An example of the breakdown
of the total numerical values of the moments
into particular contributions due to $G_i$ terms for $\mu=m_b=4.75~{\rm GeV}$,
$\al_s(M_Z)=0.118$, that are chosen as our typical values (see below), 
is given in Tables~\ref{tab2},~\ref{tab1}.

\begin{table}[htb]
\begin{center}
\begin{tabular}{|c|c|c|c|} \hline
$n$&${\cal C}_n^C$& ${\cal C}_n^1$  & ${\cal C}_n^2$ \\ \hline
10 &0.26388 & -0.02508  &0.00320  \\
12 &0.22195 & -0.02055  &0.00252  \\
14 &0.19059 & -0.01724  &0.00200  \\
16 &0.16652 & -0.01477  &0.00159  \\
18 &0.14763 & -0.01288  &0.00125  \\
20 &0.13247 & -0.01139  &0.00099  \\ \hline
\end{tabular}
\end{center}
\caption{ The 0th, 1st and 2nd order continuum
contributions to the moments}
\label{tab1}
\end{table}

From the Tables 1,2 one sees that
the expansion~(\ref{gfcor}) does
not work for the pole contribution
even at relatively large scale
($\mu=m_b$), i.e. next terms are larger than the lower order terms.
In Voloshin's paper
these  corrections were treated by a
redefinition of the scale in the leading order approximation. This
resulted in the introduction of the function $h(\gamma)$ that determines
the appropriate scale for a given moment.
Because only the first order corrections ($V_1$ from
Eq. (\ref{potcor}))
were taken into account (the contributions due to $V_2$ were
not considered)
such a redefinition of the scale in the leading order approximation 
does allowed to absorb them completely into the Coulomb Green function 
requiring $G_1=0$. 
However, this leads to an extremely low normalization point
where 
the ordinary PT for the potential 
breaks down and high order corrections become
uncontrollable (see Sect. 4 for the details).

\subsection{Optimized computation}
\noindent Now we turn to  
our method of computating the
nonrelativistic Green function with optimization of contributions due
to higher order corrections to the potential. 
In contrast to QED where the Coulomb
term exhausts the potential in the nonrelativistic approximation and
hence generates 
the full Green function, in QCD the potential itself is a
PT expansion in $\al_s$. Due to the presence of $\log(\mu r)$ terms in
this expansion the nonrelativistic Green function cannot be obtained
in  closed form and  some approximations have to be employed.
The perturbative expansion in the corrections to the potential looks
natural for this purpose. 
Now, 
for computation of the best approximation for the nonrelativistic 
GF (\ref{gfcor}) we apply the technique of optimization of the expansion 
(\ref{gfcor}) with regard to higher order corrections using the freedom
of redefinition of the leading order approximation by the proper
choice of the Coulomb coupling constant \cite{PenPiv}.
It is clear that  $\al_s$ in the $\overline{\rm MS}$ scheme 
is not necessarily the best choice for the coupling 
if  the Coulomb solution is used  as 
a leading approximation for the nonrelativistic GF.
To improve the situation  we introduce
an effective Coulomb charge by adding a part of the higher order 
constant contribution to the leading term. We parametrize this part 
with
a parameter $t$. To some extent this is reminiscent of the simple change of
the scale $\mu$, though in reality it is a more sophisticated procedure.
Its intension is to minimize (or optimize) the contribution of the
logarithmic terms in the potential that  cannot be treated exactly. 
The $\mu$ dependence is still present and  controlled by the ordinary
renormalization group invariance. 
Moreover, in contrast to ref.~\cite{Vol2} where different
normalization points are used for the analysis of the 
soft and hard corrections,
in our approach we may use a unique expansion parameter $\al_s(m_b)$ 
and the corrections related to the matching are absent.
Thus, the Coulomb part of the GF is determined  by
an effective Coulomb parameter $\al_C(t)$  
\be
\tilde V_C=-{C_F\al_C(t)\over r}
\label{Vlmod}
\ee
with $\al_C(t)$ being a function of $\al_s$
\be
\al_C(t)=\al_s\left(1+{\al_s\over 4\pi}(1-t)C^1_0\right).
\label{almod}
\ee
The explicit formula for such a decomposition
to one loop (first order approximation
for the potential) reads
\begin{eqnarray}
  \label{optim}
V(r)&=&-{C_F\al_s\over r}
\left(1+\frac{\al_s}{4\pi}(C_0^1+C_1^1\ln (r\mu))\right)\nonumber \\
&=&-{C_F\al_s\over r}
\left(1+\frac{\al_s}{4\pi}(1-t)C_0^1\right)
-{C_F\al_s\over r}\frac{\al_s}{4\pi}\left(
t C_0^1+C_1^1\ln (r\mu)\right)\nonumber \\
&\equiv &\tilde V_C+\frac{\al_s}{4\pi}\tilde V_1
\end{eqnarray}
with $C^1_i$ given in the Appendix.
Now
\[
\tilde V_C+\frac{\al_s}{4\pi}\tilde V_1 =  V_C+\frac{\al_s}{4\pi} V_1
\]
where the effective Coulomb term $\tilde V_C$ is given by 
(\ref{Vlmod}) with the effective charge (\ref{almod})
and $\tilde V_1$ is considered as a correction.
Thus we have added a part of the constant of the second order term
($V_1$)
to the Coulomb charge of the leading approximation.
Developing the expansion around the Coulomb GF
corresponding to $\tilde V_C$ we minimize corrections by choosing the
proper numerical value of the parameter $t$. At the final stage we
have to check that both expansions (for the nonrelativistic GF around
the Coulomb approximation and for the effective charge around the
$\al_s$) converge well. This happens to be the case
for the considered system. 
Otherwise one would have 
to conclude that the problem cannot be
treated perturbatively, i.e. one cannot use the PT expression for the
potential to generate the nonrelativistic approximation.
Technically the method results in substituting the combination
$C_F\al_s$ by $C_F\al_C$ in $G_C$ and $G_1$.
Then we choose $t(n)$ (or $\al_C$) such that ${\cal M}_n^1=0$
to get the best convergence of the series~(\ref{mseries}).

In the analysis we use moments with $10<n<20$. For this range of 
moments the contribution
of the continuum (not well known experimentally)
is sufficiently suppressed for the results
being practically independent of $s_0$ and, at the same time, the
nonperturbative power corrections due to the
gluonic condensate are  suppressed too\cite{Vol2}.
We use the normalization point $\mu =m_b$ and take into account the
first order $\al_s$ corrections to the polarization
function~(\ref{thmom}).

The result of the fit   
$\al_s(m_b)= 0.22\pm 0.02$ or $\al_s(M_Z)=0.118\pm 0.006$
is in a good agreement with other estimates \cite{PDG}.
The sum rules are much more sensitive to the
$b$-quark mass than to the strong coupling constant so it is instructive
to fix $\al_s(M_Z)=0.118$ to the ``world average'' value \cite{PDG}
and then to extract $m_b(n)$.
The results for $m_b(n)$ are given  in Table~\ref{tab3}, together
with the individual moments and the respective values of the 
effective charge. 

\begin{table}[htb]
\begin{center}
\begin{tabular}{|c|c|c|c|c|}\hline
$n$&$m_b({\rm GeV})$ & ${\cal P}_n$ & ${\cal C}_n$ & $\al_C/\al_s$ \\ \hline
10 &4.730 & 0.4143 & 0.3308 & 1.306         \\
12 &4.742 & 0.4723 & 0.2827 & 1.328         \\
14 &4.752 & 0.5391 & 0.2481 & 1.349         \\
16 &4.760 & 0.6028 & 0.2174 & 1.362         \\
18 &4.767 & 0.6786 & 0.1949 & 1.376         \\
20 &4.774 & 0.7642 & 0.1780 & 1.389         \\ \hline
\end{tabular}
\end{center}
\caption{The $n$ distribution of $m_b$,  ${\cal P}_n$, 
${\cal C}_n$ and $\al_C/\al_s$ for $\mu =m_b$, $\al_s(M_Z)=0.118$}
\label{tab3}
\end{table}

\noindent
Our final estimate of the $b$-quark mass is $m_b=4.75~{\rm GeV}$,
the average over the range $n=10-20$.
As we see, the pole contribution exceeds the one from the continuum 
and obviously cannot  be neglected as it was done in ref.~\cite{PichJam}.
What is the uncertainty of this result?
The total error  from the specific choice of $k(s)$ 
in (\ref{var}) and
from the  numerical value for $s_0$ is about $\pm 0.1\%$.
However, the main uncertainty originates from higher
orders of PT for the potential~(\ref{potcor})
or from the expansion of the GF in eq.~(\ref{gfcor}).
This can be estimated   from the $n$ distribution (Table~\ref{tab3})
or from the $\mu$ dependence of the
results which formally 
must be canceled by the higher order contributions
(Table~\ref{tab4}).
\begin{table}[htb]
\begin{center}
\begin{tabular}{|c|c|c|c|c|}\hline
$\mu({\rm GeV})$&$m_b({\rm GeV})$&$N_{pt}$&$\al_C/\al_s$&$\al_s(\mu)$\\ \hline
$m_b$        & 4.752 &  0.631 &   1.349  &  0.218           \\
4.0          & 4.763 &  0.608 &   1.254  &  0.231           \\
3.0          & 4.782 &  0.564 &   1.245  &  0.257           \\
2.0          & 4.799 &  0.478 &   1.118  &  0.308           \\
1.2          & 4.767 &  0.273 &   0.879  &  0.428           \\ \hline
\end{tabular}
\end{center}
\caption{The $\mu$ dependence of $m_b$,  $N_{pt}$,
$\al_C/\al_s$ and $\al_s$ for $n=14$, $\al_s(M_Z)=0.118$}
\label{tab4}
\end{table}

\noindent
Note that at the scale $\mu\sim m_b$ both the hard gluon
corrections which are parametrized by the factor
\be
N_{pt}=1-4 C_F{\al_s\over \pi}
\label{Np}
\ee
and the soft gluon corrections  parametrized by the ratio $\al_C/\al_s$
are of the same order $\sim 35\%$ so this seems to be an
optimal normalization point. For lower scale  the
hard corrections become large while for higher scale
the same is true for the soft corrections.
These latter corrections can be also estimated directly
by adding  the $G_2$ part of the  nonrelativistic GF.
The relative weight of  $\al_s$ corrections can be
found from Table~\ref{tab5}.
In the Coulomb approximation the hard gluon corrections, {\it i.e.}
the perturbative factor in eq.~(\ref{thmom}) is neglected while
the 1st order corrections include this factor and the $G_1$ part of
the  nonrelativistic GF. In the second order estimate we
take into account the $G_2$ term but do not use the optimization procedure.
Thus collecting all the data we estimate
the high order corrections to the $b$-quark mass to be about $1\%$.
Our final value is $m_b=4.75\pm 0.04$ GeV. 

\begin{table}[htb]
\begin{center}
\begin{tabular}{|l|c|}\hline
Order of approximation  &  $m_b({\rm GeV})$ \\  \hline
pure Coulomb approximation      &  4.700       \\
total 1st order $\al_s$ corrections included
&  4.752   \\
2nd order $\al_s$ corrections to the potential included
&  4.776   \\ \hline
\end{tabular}
\end{center}
\caption{The convergence of  perturbation theory for $m_b$
at $n=14$, $\mu =m_b$, $\al_s(M_Z)=0.118$}
\label{tab5}
\end{table}

Let us emphasize that we have used
the $\al_s^2$ corrections to the potential~(\ref{potcor}) only as
an estimate of the typical size of all $\al_s^2$ corrections.
The final answer in the next-to-leading order can be obtained
only after 
all $\al_s^2$ corrections (including
relativistic ones) to the polarization function in the threshold region
are available.

\section{Comparison with the previous analyses}

In ref.~\cite{Vol2} the low normalization 
point $\mu = 1~{\rm GeV}$
is used which is effectively a way of reshuffling
the PT series in higher orders.
The hard gluon correction was separated and taken at $\mu = 0.632m_b$.
Using these prescriptions we
reproduce the results of ref. \cite{Vol2}
$m_b = 4.827\pm 0.007~{\rm GeV}$,
$\al_s(M_Z) = 0.109\pm 0.001$ within our approach.

The precision of these 
results is however strongly overestimated: no variation due to higher 
order
corrections (which at this level reduces effectively
to the change of the parameter $\mu$) is allowed.
The quoted errors just reflect the 
difference between
different moments while the  $\mu$ dependence is
fixed from the very beginning. 
Also, 
although the  correction induced by using  different normalization points
for hard and soft regions respectively 
is formally of the next (second) order in
$\al_s$, they can be very essential because of the low normalization point.
This can be seen clearly  from the fact that
at $\mu\sim 1~{\rm GeV}$ the first order
hard gluon corrections  are about $80\%$ so that PT does not converge.
On the other hand we found that the second order correction
to the moments due to $V_2$ term of the potential~(\ref{potcor})
exceeds $100\%$ at  $\mu\sim 1~{\rm GeV}$ so the first order PT
can be hardly used for the analysis of the soft gluon corrections at
this scale though the first order corrections are minimized (equal to zero).
The error bars given in  ref.~\cite{Vol2} do not
account for high order corrections.
This is the reason for the discrepancy between the result
of ref.~\cite{Vol2} and LEP data on $\al_s(M_Z)$ \cite{PDG}.

Speaking of uncertainties we would like to touch the point
of using
the moments in relativistic representation against
the Laplace transform of the nonrelativistic GF.
We mentioned this problem in the beginning of Sect.~2 when explaining
details of our treatment of nonrelativistic approximation.
The parametric estimate of $O(1/n)$ for difference between moments
and Laplas transform in ref.~\cite{Vol2} is not correct for 
an arbitrary large $n$.
Let us demonstrate it in more detail now.
Two formulas for the moments
\[
(4m^2)^n\int{R(s)ds\over s^{n+1}} \qquad \mbox{vs}\qquad \int R(E)
e^{-\frac{E}{m}n}\frac{dE}{m}
\]
that are used and where
$s=(2m+E)^2$
are not equivalent up to $1/n$ terms for large $n$ for any function $R(s)$
but only for rather smooth ones.
In fact, the dominant contribution to the theoretical
value of the moments is due to Coulomb bound states
which lead to $\delta$-function singularities in $R(s)$.
So, one has to check the equality
\be
(4m_b^2)^n\int{R(s)ds\over s^{n+1}} \stackrel{?}{=}
\int R(E)e^{-\frac{E}{m_b}n}\frac{dE}{m_b}\left(1+O\left(\frac{1}{n}
\right)\right)
\label{equiv}
\ee
for $R(s)=4m_b^2\delta(s-m_C^2)$ with $m_C=2m_b(1+\Delta)$ where
$\Delta\sim\al_s^2$, $\Delta <0$.
Then one has
\[
(4m_b^2)^n\int_0^\infty{R(s)ds\over s^{n+1}} =
{(4m_b^2)^{n+1}\over (m_C^2)^{n+1}}
={1\over (1+\Delta)^{2n+2}}
\]
while the right hand side of eq.~(\ref{equiv}) is
\[
\int_0^\infty e^{-\frac{E}{m_b}n}R(E)\frac{dE}{m_b}=
{e^{-2\Delta n}\over 1+\Delta}
\]
or
\be
{1\over (1+\Delta)^{2n+1}}\qquad {\rm vs} \qquad e^{-2\Delta n}
\label{polch3}
\ee
and these two quantities are not related by the
estimate~(\ref{equiv}),
i.e. with an accuracy of $O(1/n)$
uniformly in $n$ and $\Delta$. Indeed for large $n$
and fixed negative $\Delta$ one has
\[
\left. {e^{2\Delta n}\over (1+\Delta)^{2n+1}}\right|_{n\rightarrow \infty}
=0\ne 1+O\left(\frac{1}{n}\right).
\]
They are close to each other with the stated accuracy 
only for $|\Delta|^2 n<<1$ that can be easily checked by expanding both
quantities in a Taylor series for small $|\Delta|$.
For a smooth function, however, the estimate~(\ref{equiv}) is
valid. For instance, for $R(s)=(s/4m_b^2)
\theta(s-4m_b^2)=(1+E/2m_b)^2\theta(E)$
one has
\be
(4m_b^2)^n\int_{4m_b^2}^\infty{R(s)ds\over s^{n+1}} =
(4m_b^2)^{n-1}\int_{4m_b^2}^\infty{sds\over s^{n+1}} =
{1\over n-1}
=\frac{1}{n}+\frac{1}{n^2}+\ldots
\label{conch1}
\ee
while
\be
\int_0^\infty e^{-\frac{E}{m_b}n}R(E)\frac{dE}{m_b}=
\int_0^\infty e^{-\frac{E}{m_b}n} 
\left(1+\frac{E}{m_b}+\frac{E^2}{4m_b^2}\right)\frac{dE}{m_b}
=\frac{1}{n}+\frac{1}{n^2}+\frac{1}{2n^3}
\label{conch2}
\ee
in accordance with eq.~(\ref{conch1}) and eq.~(\ref{equiv}).
This observation in fact gives an additional constraint
from above on possible $n$. 
Meanwhile in ref.~\cite{Vol2} the upper bound on allowed 
values of $n$ was connected only with the contribution
of the gluon condensate.
We, however, should stress that  in ref.~\cite{Vol2}
the region of the parameters with $|\Delta |n\sim 1 $ 
($\alpha_s^2 n\sim 1 $) was considered
where the difference between two expressions in eq.~(\ref{polch3})
can be  estimated {\it numerically} as $O(1/n)$ or $O(\alpha_s^2)$. 

The results of ref.~\cite{PichJam}
$m_b = 4.60\pm 0.02~{\rm GeV}$, $\al_s(M_Z)=0.119\pm 0.008$
are reproduced as soon as the pole term is omitted.
This is the reason for having smaller value of $m_b$.
In fact, omitting the pole contribution leads to a huge
(about $50\%$) change  of the moments (see Table~\ref{tab3})
but only to $\sim 3\%$
variation of the $b$-quark mass.
Then the way the potential is treated in ref.~\cite{PichJam}
seems not to be
correct to us. A simple use of the running coupling constant
$\al_V(q)$ with $q=\sqrt{s-4m_b^2}$
which is defined in all orders of PT as
$V(r)=-C_F\al_V(r)/r$
instead of $\al_s$ in the
leading order nonrelativistic result (Sommerfeld factor) \cite{PichJam}
is not
consistent with the expansion of the GF~(\ref{gfcor})
that is
generated by PT of the potential~(\ref{potcor})
though it does catch a part of
the contribution.
This can be directly found by comparing the data listed
in Table~\ref{tab1} and Table~\ref{tab6} which contains
the continuum
contributions ${\cal Z}_n$ to the moments computed
according to the prescription of ref. \cite{PichJam}
by inserting the running coupling $\al_V(q)$
into the Sommerfeld factor. 
Note that in our approach this
procedure corresponds to inserting
$\al_V(q)$ instead of $\al_s$ to the Coulomb GF $G_C$.
In  Table~\ref{tab6} the quantities  
${\cal Z}_n^i$ are defined as follows
\be
{\cal Z}_n^i= 
(4m_b^2)^n\int_{4m_b^2}^\infty 
{ds\over s^{n+1}}R^C(s)|_{\al_s\rightarrow\al^{(i)}_V(q)}
\label{zdef}
\ee
where $\al^{(i)}_V$ is a series for $\al_V$ in $\al_s$ up to the $i$th order
\[
\al^{(0)}_V=\al_s, \qquad \al^{(1)}_V(q)
=\al_s\left(1+\frac{\al_s}{4\pi}\left(a_1-2\beta_0\ln\left(\frac{q}{\mu}
\right)\right)\right)
,\qquad \ldots
\]
($\al^{(2)}_V(q)$ and numerical values of parameters 
can be directly found from eqs.~(\ref{potcorr1},~\ref{potcorr2}) in the
Appendix).
The quantity $R^C(s)$ for $s>4m_b^2$ reads 
\[
R^C(s)={18\pi\over m_b^2} N_{pt}~{\rm Im} G_C\left(s+i\epsilon\right) 
{=}~\frac{9}{2} N_{pt}
\left({\pi C_F \al_s\over 1-\exp(-\pi C_F \al_s/v)}
\right),
\]
$v$ being the $b$-quark velocity.
Note that in eq.~(\ref{zdef}) $\al_s$ is not substituted by $\al_V$
in the perturbative factor $N_{pt}$ (\ref{thmom},\ref{Np}).
Thus the quantities 
${\cal Z}_n^1$ and  ${\cal Z}_n^2$ should be compared with
${\cal C}_n^C+{\cal C}_n^1$
and ${\cal C}_n^C+{\cal C}_n^1+{\cal C}_n^2$ respectively.
From  Tables~\ref{tab1},~\ref{tab6} one finds that within 
the approach of ref.~\cite{PichJam} the corrections 
to the leading order continuum contribution are about 
2-4 times underestimated. 
In any case,  the main difference between  the present analysis and
that of ref.~\cite{PichJam} is just the neglect of 
the pole contribution.

\begin{table}[htb]
\begin{center}
\begin{tabular}{|c|c|c|c|}\hline
$n$& ${\cal Z}_n^0\equiv{\cal C}_n^C$ & ${\cal Z}_n^1$ & ${\cal Z}_n^2$ \\
\hline
10 & 0.26388 & 0.25534   & 0.25273  \\
12 & 0.22195 & 0.21564   & 0.21380  \\
14 & 0.19059 & 0.18582   & 0.18450  \\
16 & 0.16652 & 0.16285   & 0.16188  \\
18 & 0.14763 & 0.14475   & 0.14403  \\
20 & 0.13247 & 0.13018   & 0.12964  \\ \hline
\end{tabular}
\end{center}
\caption{The corrections to the continuum contribution computed
by inserting the running coupling $\al_V(k^2)$
into the Sommerfeld factor for $\mu =m_b$, $\al_s(M_Z)=0.118$}
\label{tab6}
\end{table}

Furthermore the results in ref.~\cite{PichJam} 
exhibit a much weaker $\mu$ dependence
and the contribution of high orders of PT is claimed to be
much smaller. The reason is that the main corrections and the strong 
$\mu$ dependence
originate from the pole contribution which is omitted in ref.~\cite{PichJam}.

\section{Conclusion.}

\noindent We reanalyzed the sum rules for $\Upsilon$ system
and obtained new numerical estimates of the strong coupling constant
and the bottom quark pole mass.
Our results are different 
from those  two existing papers on this subject
and lie outside
error bars given by the authors.
Our results and the reasons for the discrepancy are:
\begin{itemize}
\item
We found $m_b=4.75\pm 0.04~{\rm GeV}$ and $\al_s(M_Z)=0.118\pm 0.006$
consistent with  LEP data.
The results are insensitive to the choice of $k$ in eq.~(\ref{var}),
to $s_0$, and whether resumed or plain PT is used for integration beyond
$s_0$. The main theoretical uncertainty of these values
comes from the higher order  corrections which induce  
an $n$ and $\mu$ dependence of the result.
It is explicitly shown that the part of
the corrections   connected to the potential is large 
and leads to less accurate
estimates of the mass and the coupling constant than claimed
before.
Thus the main problem at the present stage of calculation is the 
consistent treatment of the next-to-leading $\al_s$ corrections. 
\item
The  pole contribution to the
theoretical spectral density is omitted in ref.~\cite{PichJam}. 
This results in the
a significant difference in the determination of 
$m_b$ (which we consider as erroneous) and leads to the
impressive stability against a variation of  
$\mu$ (or against inclusion
of higher orders of PT) because it is the pole part that is
strongly $\mu$ dependent.
\item
In the paper \cite{Vol2} the low normalization scale is used
where the validity of the expansion around Coulomb solution
due to huge high orders $\al_s$ corrections  is
questionable.
The uncertainty of the results  is strongly
underestimated in  ref.~\cite{Vol2}.
\end{itemize}

\vspace{3mm}
\noindent
{\large \bf Acknowledgements}\\[2mm]
A.A.Pivovarov acknowledges discussions
with K.G.Chetyrkin and K.Melnikov.
This work is partially supported by BMBF under contract
No. 057KA92P and by Volkswagen Foundation under contract
No.~I/73611. A.A.Pivovarov is  
supported in part by
the Russian Fund for Basic Research under contracts Nos.~96-01-01860
and 97-02-17065. A.A.Penin greatfully
acknowledges partial financial support by
the Russian Fund for Basic Research under contract
97-02-17065.

\section*{Appendix}
The corrections to the Coulomb potential read \cite{Peter}
$$
V_C(r)=-{C_F\al_s(\mu)\over r},
$$
\be
V_1(r)=V_C(r)(C_0^1+C_1^1\ln(r\mu)),
\label{potcorr1}
\ee
$$
V_2(r)=V_C(r)(C_0^2+C_1^2\ln(r\mu)+C_2^2 \ln^2(r\mu)),
$$
where
$$
C_0^1=a_1+2\beta_0\gamma_E,
$$
\[
C_0^2=\left({\pi^2\over 3}+4\gamma_E^2\right)\beta_0^2
+2(\beta_1+2\beta_0a_1)\gamma_E+a_2,
\]
$$
C_1^1=2\beta_0,
$$
$$
C_1^2=2(\beta_1+2\beta_0a_1)+8\beta_0^2\gamma_E,
$$
\be
C_2^2=4\beta_0^2,
\label{potcorr2}
\ee
$$
a_1={31\over 9}C_A-{20\over 9}T_Fn_f,
$$
\[
a_2= \left({4343\over 162}+6\pi^2-{\pi^4\over 4}
+{22\over3}\zeta(3)\right)C_A^2-
\left({1798\over 81} + {56\over 3}\zeta(3)\right)C_AT_Fn_f-
\]
$$
-\left({55\over 3} - 16\zeta(3)\right)C_FT_Fn_f
+\left({20\over 9}T_Fn_f\right)^2,
$$
$$
\beta_0={11\over 3}C_A-{4\over 3}T_Fn_f,\qquad
\beta_1={34\over 3}C_A^2-{20\over 3}C_AT_Fn_f-4C_FT_Fn_f,
$$
$$
C_A=3, \qquad C_F={4\over 3}, \qquad T_F={1\over 2},
$$
$\gamma_E=0.57\ldots$ is the Euler constant, $\zeta(x)$ is 
the Riemann $\zeta$-function, and $n_f$ is the number of light flavors.

The nonrelativistic Coulomb GF reads \cite{Braun}
\be
G_C(0,0,k)=-{C_F\al_sm_b^2\over 4\pi}\left({k\over C_F\al_sm_b}+
\ln\left({2k\over C_F\al_sm_b}\right)
-\sum_{m=0}^\infty F(m)\right)
\label{G0}
\ee
where
\be
F(m)={1\over m+1-{\displaystyle {C_F\al_s m_b\over 2k}}} -{1\over
m+1}. 
\label{fdef}
\ee 
Evaluation of the corrections to the  Coulomb GF 
is straitforward but rather complicated. In the calculation
we used the representation
of the Coulomb GF $G_C(x,y,k)$ as a sum of Laguerre polynomials \cite{Vol1}.
The result reads
$$
G_1(0,0,k)={\al_s\over 4\pi}{C_F\al_sm_b^2\over 4\pi}\left(
\sum_{m=0}^\infty F^2(m)(m+1)
\left(C_0^1-\ln\left({2k\over\mu}\right)C_1^1+\Psi_1(m+2)C_1^1\right)
-\right.
$$
$$
2\sum_{m=1}^\infty\sum_{n=0}^{m-1}
F(m)F(n)
{n+1\over m-n}C_1^1 +
$$
\be
2\sum_{m=0}^\infty F(m)
\left(\left(C_0^1-\ln\left({2k\over\mu}\right)C_1^1\right)-
(2\gamma_E+\Psi_1(m+1))C_1^1\right) -
\label{G1}
\ee
$$
\left.C_0^1\ln\left({2k\over\mu}\right)+
 \left(\gamma_E\ln\left({2k\over\mu}\right)+
{1\over 2}\ln^2\left({2k\over\mu}\right)
\right)C_1^1
\right),
$$
\[
G_2(0,0,k)=\left({\al_s\over 4\pi}\right)^2{C_F\al_sm_b^2\over
4\pi}\left( \sum_{m=0}^\infty F^2(m)
\left((m+1)\left(C_0^2-\ln\left({2k\over\mu}\right)C_1^2+
\right.\right.\right.
\]
$$
\left.\left.
\ln^2\left({2k\over\mu}\right)C_2^2\right)+
(m+1)\Psi_1(m+2)\left(C_1^2-2\ln\left({2k\over\mu}\right)
C_2^2\right)+I(m)C_2^2\right) +
$$
\be
2\sum_{m=1}^\infty\sum_{n=0}^{m-1}
F(m)F(n)\left(-
{n+1\over m-n}\left(C_1^2 -2\ln\left({2k\over\mu}\right)C_2^2\right)
+J(m,n)C_2^2\right)+
\label{G2}
\ee
$$
2\sum_{m=0}^\infty F(m)
\left(\left(C_0^2-\ln\left({2k\over\mu}\right)C_1^2+
\ln^2\left({2k\over\mu}\right)C_2^2\right)-\right.
$$
\[
\left.(2\gamma_E+
\Psi_1(m+1))
\left(C_1^2 -2\ln\left({2k\over\mu}\right)C_2^2\right) +K(m)C_2^2\right) -
\]
$$
\left.C_0^2\ln\left({2k\over\mu}\right)+
\left(\gamma_E\ln\left({2k\over\mu}\right)+
{1\over 2}\ln^2\left({2k\over\mu}\right)\right)
C_1^2 
+L(k)C_2^2\right)
$$
where
\[
I(m)=(m+1)\left(\Psi^2_1(m+2)-\Psi_2(m+2)+{\pi^2\over3}
-{2\over(m+1)^2}\right)-
\]
$$
2(\Psi_1(m+1)+\gamma_E),
$$
$$
J(m,n)= 2{n+1\over m-n}\left(\Psi_1(m-n)-{1\over n+1}+2\gamma_E\right)+
$$
\[
2{m+1\over m-n}(\Psi_1(m-n+1)-\Psi_1(m+1)),
\]
$$
K(m)=2(\Psi_1(m+1)+\gamma_E)^2+\Psi_2(m+1)-\Psi_1^2(m+1)+2\gamma_E^2,
$$
\[
L(k)=-\left(\gamma_E+{\pi^2\over 6}\right)\ln\left({2k\over \mu}\right)
-\gamma_E\ln^2\left({2k\over \mu}\right)-{1\over 3}
\ln^3\left({2k\over \mu}\right),
\]
$$
\Psi_1(x)={\Gamma'(x)\over\Gamma(x)},
\qquad\Psi_2(x)=\Psi_1'(x)
$$
and $\Gamma(x)$ is the Euler $\Gamma$-function.

\begin{thebibliography}{99}

\bibitem{NSVZ} V.A.Novikov {\it et al.}, Phys.Rev.Lett. {\bf 38}(1977)626;\\
               V.A.Novikov {\it et al.}, Phys.Rep. {\bf C41}(1978)1.
\bibitem{KrasnPiv}N.V.Krasnikov and A.A.Pivovarov, 
Phys.Lett. {\bf 112B}(1982)397.
\bibitem{Vol1} M.B.Voloshin, Yad.Fiz. {\bf 36}(1982)247;\\
               M.B.Voloshin and Yu.M.Zaitsev, Usp.Fiz.Nauk
               {\bf 152}(1987)361.
\bibitem{Chet} K.G.Chetyrkin, J.H.K\"uhn, M.Steinhauser,
               Nucl.Phys. {\bf B482}(1996)213.
\bibitem{Vol2} M.Voloshin, Int.J.Mod.Phys. {\bf A10}(1995)2865.
\bibitem{PichJam} M.Jamin, A.Pich, Nucl.Phys. {\bf B507}(1997)334.


\bibitem{PDG} Particle Data Groop, Phys.Rev. {\bf D54}(1996)1.
\bibitem{Braun} M.A.Braun, ZhETP Lett. {\bf 27}(1968)652;\\
                R.Barbieri, P.Christillin and E.Remiddi, Phys.Rev.
                {\bf A8}(1973)2266.

\bibitem{Peter}M.Peter, {Phys.Rev.Lett.} {\bf 78}(1997)602.
\bibitem{BethSal} E.E.Salpeter and H.A.Bethe, Phys.Rev.
                  {\bf 84}(1951)1232;\\
                  R.Barbieri and E.Remiddi, Nucl.Phys.
                  {\bf B141}(1978)413.

\bibitem{PenPiv}  A.A.Penin and  A.A.Pivovarov, Phys.Lett.
                  {\bf B367}(1996)342.

\end{thebibliography}
\end{document}